\documentclass[useAMS,usenatbib]{mn2e}

\title[Variability in quasars]{Studying X-ray reprocessing and continuum variability in quasars: 
PG 1211+143}

\author[Bachev et al.]
	{R. Bachev$^{1}$, D. Grupe$^{2}$, S. Boeva$^{1}$, E. Ovcharov$^{3}$, 
	A. Valcheva$^{1, 3}$, E. Semkov$^{1}$, 
\newauthor 
	Ts. Georgiev$^{1}$,  L.C. Gallo$^{4}$ \\ 
$^{1}$Institute of Astronomy, Bulgarian Academy of Sciences, Sofia 1784, 
Bulgaria; bachevr@astro.bas.bg \\
$^{2}$Department of Astronomy and Astrophysics, Pennsylvania State University, 525 Davey Lab,
University Park, PA 16802, USA; \\ grupe@astro.psu.edu \\
$^{3}$Department of Astronomy, Sofia University, 5 James Bourchier Street, 1164 Sofia, Bulgaria \\
$^{4}$Department of Astronomy and Physics, Saint Mary's University, Halifax, NS B3H 3C3,
Canada, lgallo@ap.stmarys.ca}
\date{}

\pagerange{\pageref{firstpage}--\pageref{lastpage}} 
\pubyear{2007}

\begin{document}

\newcommand{\swift}{{\it Swift}}
\newcommand{\suzaku}{{\it Suzaku}}
\newcommand{\xmm}{{\it XMM-Newton}}
\newcommand{\chandra}{{\it Chandra}}
\newcommand{\ax}{$\alpha_{\rm X}$}
\newcommand{\rb}[1]{\raisebox{1.5ex}[-1.5ex]{#1}}
\newcommand{\msun}{$M_{\odot}$}
\newcommand{\dM}{\dot M}
\newcommand{\dMM}{$\dot{M}/M$}
\newcommand{\dMedd}{\dot M_{\rm Edd}}
\newcommand{\plm}{$\pm$}
\newcommand{\nh}{$N_{\rm H}$}
\newcommand{\auv}{$\alpha_{\rm UV}$}
\newcommand{\aox}{$\alpha_{\rm ox}$}

\maketitle

\label{firstpage}

\begin{abstract}

We present the results from a monitoring campaign of the Narrow-Line Seyfert~1 galaxy PG~1211+143. The object was 
monitored with ground-based facilities (UBVRI photometry; from February to July, 2007) and with \swift\ 
(X-ray photometry/spectroscopy and UV/Optical photometry; between March and May, 2007).
 We found PG~1211+143 in a historical 
low X-ray flux state at the beginning of the \swift\ monitoring campaign in March 2007.
It is seen from the light curves that 
while violently variable in X-rays, the quasar shows  little variations in optical/UV bands. 
The X-ray spectrum in the low state is similar to other Narrow-Line Seyfert 1 galaxies during their low-states  
and can be explained by a strong partial covering absorber or by X-ray reflection onto the disk. With the current data
set, however, it is not possible to distinguish between both scenarios. 
The interband cross-correlation functions indicate a possible reprocessing of the X-rays into the 
longer wavelengths, consistent with the idea of a thin accretion disk, powering the quasar. 
The time lags between the X-ray and the optical/UV light curves, ranging from $\sim$2 to $\sim$18 days for the different 
wavebands, scale approximately as $\sim \lambda^{4/3}$, but appear to be somewhat larger than expected for this object, 
taking into account its accretion disk parameters. Possible implications for the location of the X-ray 
irradiating source are discussed.

\end{abstract}


\section{Introduction}

Powered by accretion, supposedly onto a supermassive black hole, quasars (Active Galactic Nuclei, AGN) are long 
known mostly as highly energetic, exotic objects in the hearts of the galaxies. Not until recently was their 
key role in galaxy evolution realized, revealed mostly as a strong correlation between the properties of 
the central black hole and that of the host galaxy (Magorrian et al. 1998, Ferrarese \& Merritt 2000). 
Studying quasars, therefore, is not only important to understand the underlying physics; it can also help 
to shed some light on the strange interplay between the accreting matter from the host and outflows from 
the center, which ultimately shape both -- the black hole and the galaxy.

Although a general picture of the structure of a typical quasar seems to be widely accepted 
(e.g. Elvis 2000; see also Krolik 1999), there are still many details in this picture that are not 
fully understood. Many of the problems to be solved concern AGN continuum variability -- a rather common property, 
observed in practically all energy bands. Its universality indicates perhaps that variability should be 
an intrinsic property of the processes, responsible for continuum generation. The optical/UV to X-ray part 
of the continuum spectrum, as typically assumed, originates from an accretion disk around the central 
supermassive black hole. 

Generally, X-ray variability can be caused by several factors: a change in the accretion rate; 
variable absorption (e.g. Abrassart \& Czerny 2000); variable reflection (e.g. through a change 
of the height of the irradiating source, Miniutti \& Fabian 2004; see also Gallo 2006, Done \& Nayakshin 2007); 
some combination of reflection and absorption (e.g. Chevalier et al. 2006; Turner \& Miller 2009); 
hot spots orbiting the central black hole (Turner et al. 2006; Turner \& Miller 2009); 
local flares (Czerny et al. 2004), etc.

The AGN type with the strongest X-ray variability is the class of Narrow-Line Seyfert 1 galaxies 
(NLS1s, e.g. Osterbrock \& Pogge, 1985). In addition, NLS1s show the steepest X-ray spectra  
seen among all AGN (e.g. Boller et al. 1996, Brandt et al. 1997,
Leighly 1999a, b, Grupe et al. 2001). Most of their observed
properties, like spectral slopes, FeII and [OIII] line ratios, CIV shifts, etc., appear to be driven by the 
relatively high Eddington ratio $L/L_{\rm Edd}$ in these objects (e.g. Sulentic et al. 2000, Boroson 2002, 
Grupe 2004, Bachev et al. 2004).  

What concerns the optical/UV variability, the picture there is even more puzzling. There are many 
factors that can contribute to the variations of the optical flux, but most of them can account for the 
long-term (months to years) changes. There are often reported in many objects, however, short-term (day to week) 
optical/UV variations, simultaneous with or shortly lagging behind the X-ray variations. 
An interesting idea that can explain such a behaviour is reprocessing of the highly variable X-ray 
emission into optical/UV bands.

In this paper we address the question of the relations between the X-ray and the optical/UV emission by studying the
variability from X-rays to I-band of the NLS1 PG 1211+143. This NLS1 has been the target of almost all 
major X-ray observatories since EINSTEIN (Elvis et al. 1985). The X-ray continuum displays a strong and variable 
soft X-ray excess (Pounds \& Reeves 2007). From XMM-Newton RGS data, Pounds et al. (2003) suggested the 
presence of high-velocity outflows in PG~1211+143, a result that was questioned by Kaspi \& Behar (2006). 
However, high-velocity outflows seen in X-rays have been repeatedly reported (e.g. Leighly et al. 1997) and 
new XMM-Newton data of PG 1211+143 (Pounds \& Page 2006) seem to confirm the previous claims made by Pounds et al. (2003).
 
Our primary goal is to find out if and how the X-ray variations are transferred into the longer-wavelength continuum. 
Time delays between the flaring X-ray emission, presumably coming from a compact, central source and the optical/UV 
light curves are expected, provided the X-rays are reprocessed in the outer, colder part of an accretion disk.  
Such a study may have implications on two important problems -- the radial temperature distribution of an accretion disk 
(and hence -- the type of the disk) and the location of the X-ray source, based on how much the disk "sees" it. 

This paper is organized as follows: In Section 2 we describe the \swift\ and ground-based optical monitoring 
observations. Section 3 focuses on presenting the results of this study and in Section 4 we discuss these results in the
context of the general picture of AGN. Throughout the paper spectral indexes are denoted as energy spectral indexes
with $F_{\nu} \propto \nu^{-\alpha}$. Luminosities are calculated assuming a $\Lambda$CDM cosmology with $\Omega_{\rm M}$=0.27, 
$\Omega_{\Lambda}$=0.73 and a Hubble constant of $H_0$=75 km s$^{-1}$ Mpc$^{-1}$.

\section{Observations and reductions}

\subsection{Swift data}

The \swift\ Gamma-Ray Burst (GRB) explorer mission (Gehrels et al. 2004) monitored PG 1211+143 between 2007 March 08 
and May 20. Note, that scheduled observations were twice bumped by detections of Gamma-Ray-Bursts\footnote{Although 
\swift\ has turned into a multi-purpose observatory, its main focus is still on observing GRBs and therefore GRBs will 
supersede scheduled ToO observations.}, explaining the absence of segments 15 and 20 (Table A1).  
After our monitoring campaign in 2007, PG 1211+143 was re-observed by \swift\ in February 2008 (segment 24)
However, 
this observation was used to slew between two targets. Therefore, this observation is very short (188s) and no 
X-ray spectra or UVOT photometry data were obtained. This observation only allows to measure a count rate.
A summary of all \swift\ observations is given in Table\,\ref{obs_log}. The \swift\ X-Ray Telescope 
(XRT; Burrows et al. 2005)
was operating in photon counting mode (PC mode; Hill et al. 2004) and the data were reduced by the task 
{\it xrtpipeline} version 0.10.4, which is included in the HEASOFT package 6.1. 
Source photons were selected in a circular region  with a radius of 47$^{''}$ and background region of a close by 
source-free region with $r=188^{''}$. Photons were selected with grades 0--12. 
The photons were extracted with {\it XSELECT} version 2.4. The spectral data were re-binned by using 
{\it grppha} version 3.0.0 having 20 photons per bin. The spectra were analyzed with {\it XSPEC} version 12.3.1 
(Arnaud 1996). 
The ancillary response function files (arfs) were created by {\it xrtmkarf} and corrected for vignetting
and bad columns/pixels
using the exposure maps. We used the 
standard response matrix {\it swxpc0to12s0\_20010101v010.rmf}. 
Especially during the low-state the number of photons during one segment is too small to derive a spectrum with 
decent signal-to-noise. Therefore we co-added the data of several segments to obtain source and background spectra. 
In order to examine spectral changes at different flux/count rate levels, we created spectral for the low, 
intermediate, and high states with count rates CR $<$0.12 counts s$^{-1}$, 0.13$<$CR$<$0.2, and 
CR $>$0.2 counts s$^{-1}$.
This lead to high state source and background spectra co-adding the data
from 2007 April 22, May 09 and 14 (segments 018, 021, and 022),  2007 March 26 and April 02
(segments 13 and 14) for the intermediate state, and all other for the low state. 
As for the arfs, we created an arf for each segment and coadded them by
using the ftool {\it addarf} weighted by the exposure times. Due to the low number of photons in the February 2008
observation (segment 024) no spectra could be derived. Fluxes in the 0.2--2.0 and 2--10 keV band for this segment were
determined from the count rates in these bands by comparing the fluxes during the high state during segments 018,
021, and 022, assuming no spectral changes.
All spectral fits were performed in the observed 0.3--10.0 keV energy band.
In order to compare the observations from different missions we use the HEASARC tool {\it PIMMS} version 3.8. 

Data were also taken with the UV/Optical Telescope (UVOT; Roming et al. 2005), which operates between 1700--6500 \AA\,
using 6 photometry filters. Before analyzing the data, the exposures of each segment
were co-added by the UVOT task {\it uvotimsum}. 
Source counts were selected with the standard 5$^{''}$ radius in all filters (Poole et al. 2008) and background
counts in a source-free region with a radius r=20$^{''}$.
The data were analyzed with the UVOT software tool {\it uvotsource} assuming a GRB like power law 
continuum spectrum. 
The magnitudes were all corrected for Galactic reddening  
$E_{\rm B-V}=0.035$ given by Schlegel et al. (1998) using the extinction correction in the UVOT bands given in Roming et
al. (2009).

\subsection{\xmm\ data analysis}
In order to compare the results derived from the \swift\ observations we also analyzed the \xmm\ data of
PG 1211+143. \xmm\ observed PG 1211+143 on 2001 June 15 and 2004 June 21 for 53 and 57 ks, respectively
(Pounds \& Reeves 2007). Because our paper focuses on the \swift\ and ground-based monitoring campaigns in
2007 we reduced only the \xmm\ EPIC pn data. A complete analysis of these \xmm\ data sets can be found in
Pounds et al. (2003), Pounds \& Page (2006) and Pounds \& Reeves (2007). The \xmm\ EPIC pn data were reduced
in the standard way as described e.g in Grupe et al. (2004).

\subsection{Ground-based observations}

Additional broad-band monitoring in UBVRI bands was performed on 3 telescopes: the 2-m RCC and the 50/70-cm Schmidt 
telescopes of Rozhen National Observatory, Bulgaria and the 60-cm telescope of Belogradchik Observatory, Bulgaria. 
All telescopes are equipped with CCD cameras: the 2-m telescope with a VersArray CCD, while the smaller telescopes -- with SBIG ST-8. 
Identical (U)BVR$_{\rm c}$I$_{\rm c}$ filters are used in all telescopes. The ground-based monitoring covered a period of about 5 months 
(February -- July, 2007), during which the object was observed in more than 40 epochs in BVRI bands, and occasionally -- in U. 
The photometric errors varied significantly depending on the telescope, the filter, the camera in use and the atmospheric 
conditions, but were typically 0.02 -- 0.03 mag. (rarely up to $\sim$0.1 mag in some filters for the smaller instruments).

\section {Results}

\subsection{Long-term X-ray Light Curve}
\begin{figure}
 \mbox{} \vspace{9cm} \includegraphics{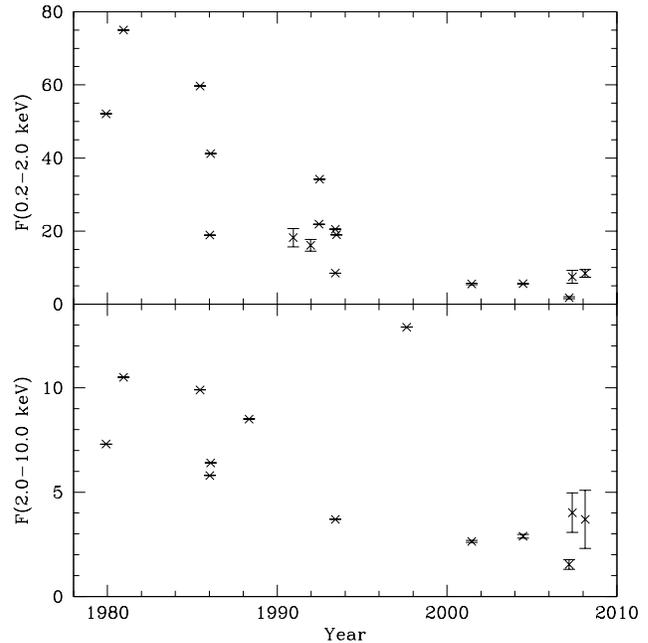} 

\caption[]{Long-term 0.2--2.0 and 2.0--10.0 keV X-ray light curves of PG 1211+143. The fluxes are given in units of 
 $10^{-12}$ ergs s$^{-1}$ cm$^{-2}$. \label{longterm}}

\label{pg1211_xray_lt_lc}
\end{figure}

Figure\,\ref{pg1211_xray_lt_lc} displays the long-term 0.2--2.0 and 2.0--10.0 keV light curves. 
Most of the data prior 2000 were taken from Janiuk et al. (2001). 
The ROSAT All-Sky Survey point at 1990.9 was taken from Grupe et al. (2001). 
The XMM 2001 and 2004 and the \swift\ fluxes were from our analysis as presented in this paper. 
PG 1211+143 has become fainter over the last decades
in both soft and hard bands with the strongest changes in the soft band. Historically, in the early 1980s,
 PG 1211+143 had been in a much brighter X-ray 
 state than over the last decade.
During the beginning of the 
 \swift\ monitoring campaign in March 2007, PG 1211+143 appeared to be in the lowest state seen so far.  
At the end of our monitoring campaign in May 2007 PG 1211+143 was back in 
the high state that was previously known from the XMM observations. The latest data point is from February 2008. 
 The 0.2--2.0 and 2--10 keV fluxes during that observation are comparable
 with the XMM observations.  As for the \swift\ data taken in 2007 and 2008 we used the flux values obtained from the
 low and high state spectra. As for the February 2008 data flux points we converted the count rates into fluxes 
 assuming an X-ray spectrum as seen during the \swift\ high states.

\subsection{\swift\ XRT and UVOT light curves}
\begin{figure}

\mbox{} \vspace{10.0cm}  \includegraphics{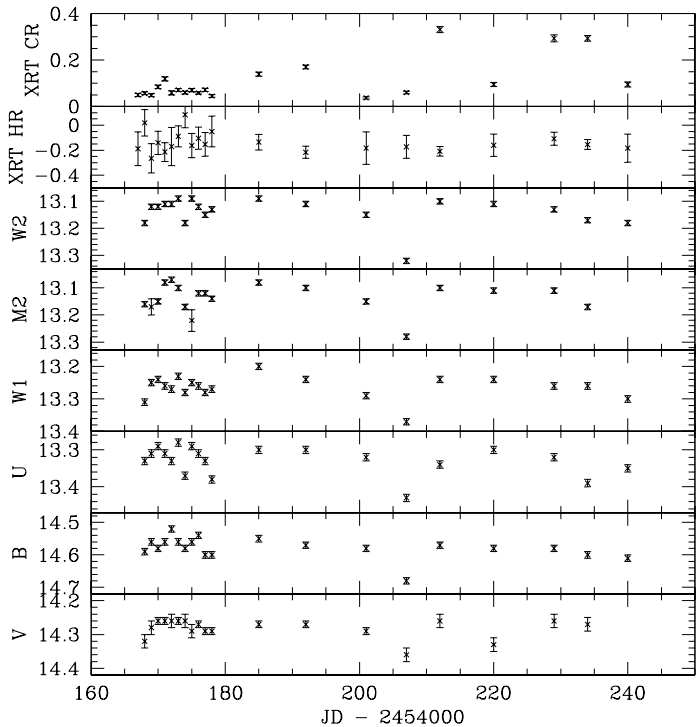} 

\caption[]{\label{swift_lc}
\swift\ XRT and UVOT light curves. The XRT count rates are given in units of counts s$^{-1}$. The
hardness ratio is defined in Section 3.2. All UVOT magnitudes are corrected for Galactic
reddening ($E_{\rm B-V}$=0.035; Schlegel et al. 1998).
}

\end{figure}

The \swift\ XRT count rates and hardness ratios\footnote{The XRT hardness ratio is defines as HR=(H-S)/(H+S)
with S and H are the counts in the 0.3--1.0 keV and 1.0--10.0 keV bands, respectively.}, 
and UVOT magnitudes are listed in Table\,\ref{xrt_uvot_res}. These values are plotted in Figure\,\ref{swift_lc}. 
At the beginning of the \swift\ monitoring campaign in March 2007, PG 1211+143 was found in a very low state. 
Compared to previous ROSAT and \xmm\ observations, reported by e.g. Grupe et al. (2001) 
and Pounds et al. (2003, 2006, 2007), 
PG 1211+143 appeared to be fainter by a factor of about 10. At the end of the monitoring campaign in May 2007, 
PG 1211+143 reached a flux level that was expected from the previous ROSAT and \xmm\ observations.
A later observation by \swift\ on 2008 February 17  found it at a level of 
0.375\plm0.045 counts s$^{-1}$ and confirmed that it returned back in a high state. The low-state, found
in March 2007, seems to be just a short temporary event. A behaviour like this is not unseen in AGN and has been 
recently reported for the NLS1 Mkn 335 by Grupe et al. (2007b, 2008a), which had been found in a historical low 
state by \swift\ and \xmm.  

Besides the X-ray variability, PG 1211+143 also displays some variability at optical/UV
wavelengths, although on a much smaller level than in X-rays. Table\,\ref{xrt_uvot_res} lists the
magnitudes measured in all 6 UVOT filters. All 6 light curves are also plotted in
Figure\,\ref{swift_lc}. The most significant drop occurred in all 6 filters during the
2007 April 17  observation. During the next observation on 2007 April 22, PG 1211+143 not only
became brighter again in the optical/UV, but also showed an increase in count rate by a factor of
almost 4 in X-rays.

\subsection{Ground-based monitoring}

\subsubsection{Secondary standards}

In order to facilitate future photometric studies of PG 1211+143, we calibrated secondary standards in the field of the object, 
shown as stars "A" (USNO B1 1039-0200330) and "B" (USNO B1 1040-0199800) in Figure 3. The magnitudes with the errors 
(due primarily to the errors of the calibration) are given in Table 1. Mostly Landolt standard sequences (Landolt 1992) 
were used for the calibration (PG 1633+099) and in some occasions -- M67 (Chevalier \& Ilovaisky 1991). 

\begin{table}
  \centering
  \caption{Field standards}
  \begin{tabular}{ccccc}
  \hline
  \hline

Star &  B & V & R & I \\

\hline

A  &  11.80 (0.08)& 11.35 (0.05)& 10.97 (0.06) & 10.71 (0.05) \\%
B  & --          &  15.34 (0.10)& 14.80 (0.07) & 14.35 (0.07) \\%

\hline
\hline
\end{tabular}
\end{table}

\begin{figure}

 \mbox{} \vspace{8.5cm} \includegraphics{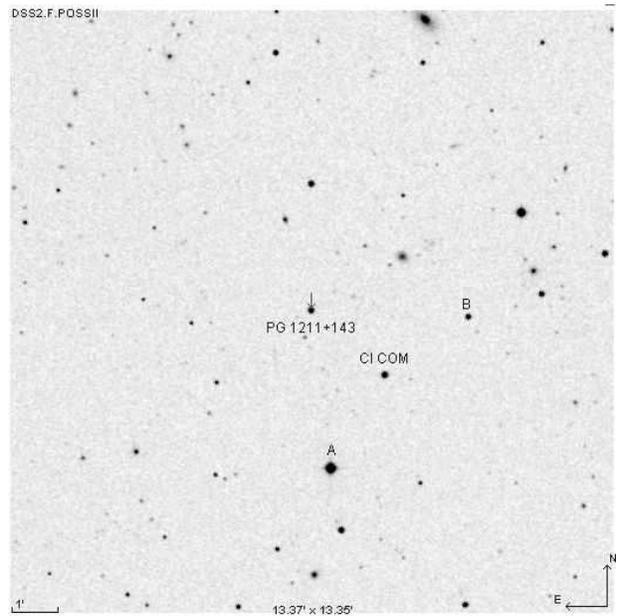} 

\caption[]{Stars in the filed of PG 1211+143, used for the ground-based differential photometry. Their optical (BVRI) magnitudes are 
calibrated and shown in Tab. 1. A nearby RR Lyr variable is also shown. CDS Aladin service was used to generate the picture.}

\end{figure}

\begin{figure*}

 \mbox{} \vspace{11.5cm} \includegraphics{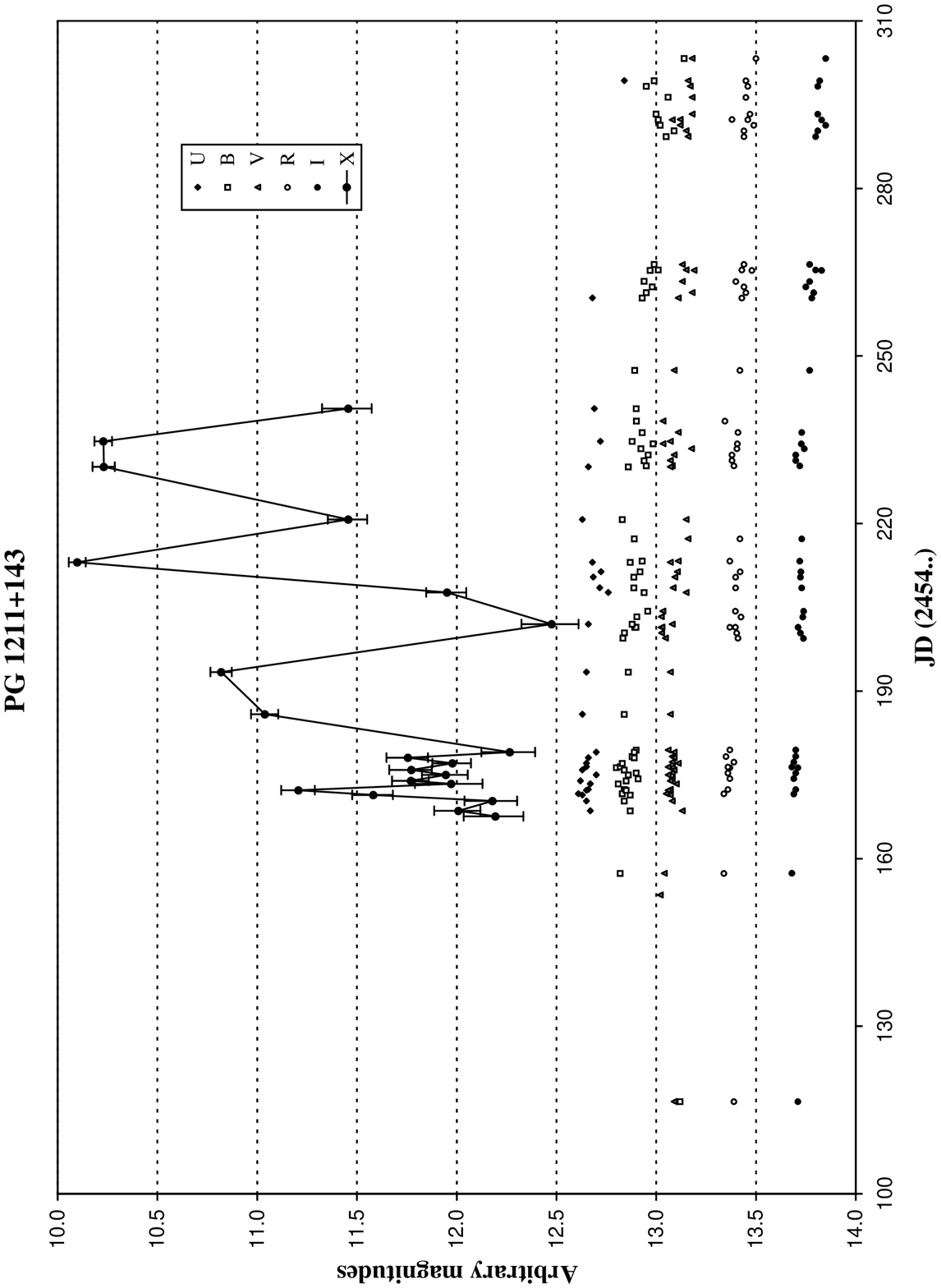} 

\caption[]{Optical (UBVRI) and X-ray light curves for the time of the monitoring (February -- July, 2007). Where necessary, arbitrary 
constants are added to the magnitudes to separate the curves and improve the presentation (see the text). The X-ray count rates are 
converted into arbitrary magnitudes (using $m_{\rm X}=-2.5\log(C.R.)+constant$) for a better comparison with the optical variability. 
The optical data are combined from all instruments.}

\end{figure*}

\subsubsection{Magnitude adjustments}

The light curve (LC) of PG 1211+143 (Sect. 3.4) is built by measuring its differential magnitude in respect to the adjacent field stars, 
none of which showed signs of variability (with the exception of a known RR Lyr type star, CI Com, located very close to the quasar).  
Since the data are collected on different instruments using different cameras (even with identical filters) it is not unusual 
for an object with strong emission lines to show a differential magnitude, slightly depending on the instrument. The reason 
for this complication is mostly related to the nature of the quasar spectrum: if a strong emission line falls in a wavelength 
region where the cameras have different sensitivities, or the filter transparency is slightly different, one may get a broad-band 
differential magnitude depending on the instrument. In our case all R-band magnitudes of the quasar had to be adjusted by 0.1 mag 
for one of the instruments (the 50/70cm Schmidt telescope), probably due to the reasons described above. 
The adjustment corrections were easily obtained through comparison of the light curves, which cover each other on many occasions. 
We should note, however, that the exact quasar magnitudes are not of importance for this study, as the variations are only considered.

Additionally, the ground-based UBV magnitudes were similarly adjusted to match the corresponding Swift magnitudes. 
Actually, after the correction for the Galactic reddening, the adjustments for the ground-based B and V magnitudes were 
very minor, typically less than 0.03 mag, which is smaller than the uncertainties of the calibrated field standards (Sect. 3.3.1).

A log of the ground-based observations, including the obtained UBVRI magnitudes of the quasar after the corrections for 
the Galactic reddening is presented in Table A3.

\subsection{Combined light curves}

The combined continuum light curves of PG 1211+143 for the time of monitoring are presented in Figs. 4 and 5. Fig. 4 compares 
the X-ray with the optical (UBVRI) variations, all transformed into arbitrary magnitudes for presentation purposes. 
The optical data are combined from all participating instruments. One sees that the erratic X-ray variations (almost 
2.5 magnitudes) hardly influence the optical flux, which shows only minor variations on a generally decaying trend. 
Figure 5 presents the most intense period of the monitoring, comparing X-ray and V-band magnitudes. 
The V-band LC for that period stays remarkably stable, with a RMS smaller or comparable to the typical 
photometric errors.

\subsection{Time delays}

In order to study the time delay dependence of the wavelength we performed a linear-interpolation cross-correlation analysis 
(Gaskell \& Sparke  1986) between the X-ray and the other bands' light curves. The interpolation between the photometric points is needed due 
to unevenly sampled data and is one of the frequently used methods. Other methods applied in the literature do not seem to obtain significantly 
different results (e.g. discrete CCF, Edelson \& Krolik 1986; z-transformed CCF, Alexander 1997) when compared. 

Figure 6 shows the interpolation cross-correlation functions, ICCF($\tau$), between the X-ray LC and the other band LCs. A maximum of an 
ICCF($\tau$) for a positive $\tau$ indicates a time delay behind the X-ray changes and is a signature of a possible reprocessing. 
Although the ICCFs are mostly negative, due to the different overall trends of the X-ray and optical/UV wave-bands, a clear maximum 
for a positive $\tau$ is evident for most wave-bands. 


Since the X-ray points distribution was far from a Gaussian, even on a magnitude scale, a rank correlation was attempted, 
but the resulting ICCFs appeared to be very similar.

Table 2 and Figure 7 show the wavelength dependence of the time lag. The wavelengths are taken from the corresponding transmission
curve of the filters used (with uncertainties associated with the band widths) and the time delays are from the ICCF maxima. 

One sees that for the I-band the highest peak of the ICCF (Figure 6) is at $\tau \simeq -2$ days, indicating a possible short 
lag of the X-rays behind the near-infrared emission. Another, lower peak at $\tau \simeq +18$ days is also evident. 
This maximum could probably be associated with reprocessing and is plotted in Figure 7 mostly to demonstrate its 
consistence with the fit (see below). However, the I-band response to the X-ray variations seems to be more complicated 
than the simple reprocessing model suggests, as we discuss later in Sect. 4


Uncertainties of the maximum of the ICCFs are difficult to assess. Although there are methods, described in the literature 
(Gaskell \& Peterson 1987), one can hardly rely completely on so computed uncertainties, since the true behaviour of the light curve 
at the places where it is interpolated is anyway impossible to predict. That is why we accepted the width of ICCF profile at an 
appropriate level around the peak as an indicative of the uncertainty. This approach is very simple and in addition incorporates 
into the errors such unknowns as the inclination of the disk in respect to the observer, the spatial size of the irradiating source, etc.
See Bachev (2009) for more discussions on these issues.

A clear relation between $\tau$ and $\lambda$ is seen and an acceptable (nonlinear) fit to the data is 
$\tau_{\lambda}\simeq 9\lambda_{\rm 5000}^{4/3}$ [days]\footnote{All the calculations here are performed using the observer's 
frame measurements. Due to the similar way the times and the wavelengths are affected by the redshift, for the quasar rest frame 
the delay in the $\tau - \lambda$ dependence {\it increases} only by $(1+z)^{-1/3}$, Sect. 4, which is only $\sim 3\%$ and is 
much less than the expected errors.}, where $\lambda_{\rm 5000}$ is $\lambda / 5000$\AA. 
Section 4 discusses possible implications of this result and how it fits 
into the model of reprocessing from a thin accretion disk.

\begin{table}
  \centering
  \caption{Wavelength dependence of the time delays with uncertainties. 
  For I-band, the highest and the first positive (in parentheses) ICCF maxima are shown (see the text).}
  \begin{tabular}{ccccc}
  \hline
  \hline

Filter &  $\lambda_{\rm 0}$  (\AA) & FWHM (\AA) & $\tau$ (days) & $\Delta \tau$  (days) \\

\hline

UVW2	& 1928         &  657           &  2.5    &	3  	   \\
UVM2	& 2246	       &  498	      &  2.5    &	3.5	   \\
UVW1	& 2600	       &  693	      &  2	    &	3  	   \\
U	& 3465	       &  785	      &  4	    &	3.5	   \\
B	& 4392	       &  975	      &  5	    &	3.5	   \\
V	& 5468	       &  769	      &  13	    &	3  	   \\
R	& $\sim$6500	       &  $\sim$700	      &  14	    &	4  	   \\
I	& $\sim$8300	       &  $\sim$1000	      &  $-$2 (18)	    &	3 (4)  	   \\

\hline
\hline
\end{tabular}
\end{table}

\begin{figure}
 \mbox{} \vspace{6.5cm} \includegraphics{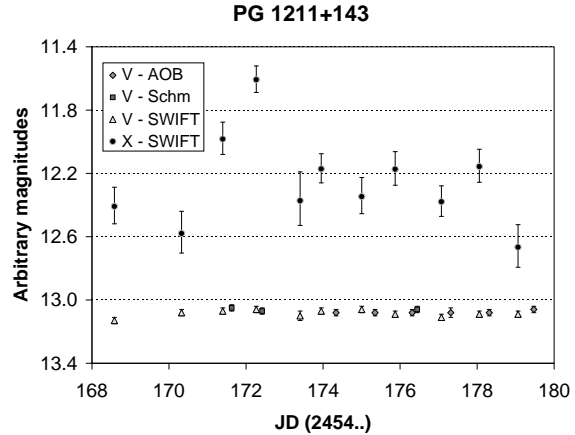} 

\caption[]{The most intensive twelve days of the monitoring, where the quasar was observed typically twice a day in the optical. 
The V-band light curve (magnitudes from the participating ground-based observatories: Rozhen Schmidt camera and Belogradchik reflector, 
and \swift\ UVOT are shown separately) and the X-ray light curve (in arbitrary magnitudes) are compared in the figure. One sees the significant 
X-rays variations and almost constant optical flux (the standard deviation of the V-band magnitudes for this 
period is smaller than the average photometric error).}

\end{figure}

\begin{figure}
 \mbox{} \vspace{8.5cm} \includegraphics{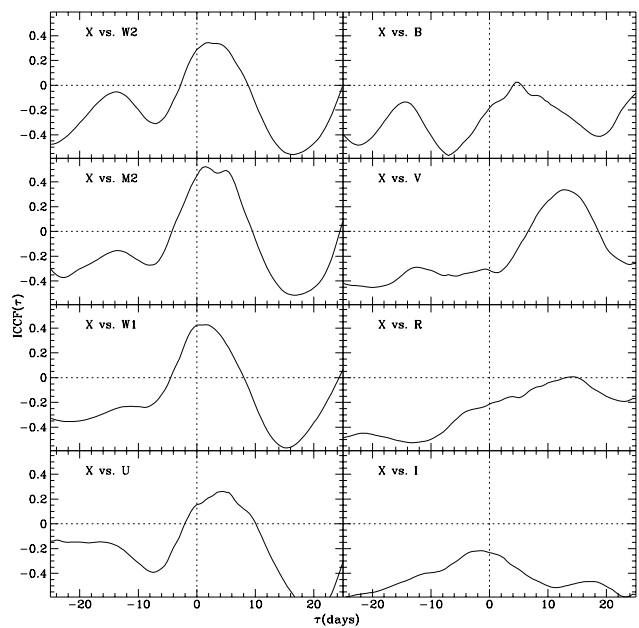} 

\caption[]{The interpolated cross-correlation functions between the X-ray light curve and the optical/UV light curves. 
A clear peak for a positive $\tau$, indicating an optical/UV delay behind the X-rays is visible in most cases. 
The ICCFs are mostly negative due to the different overall trends of the X-ray and the optical emission (see the text).}

\end{figure}

\begin{figure}
 \mbox{} \vspace{6.5cm} \includegraphics{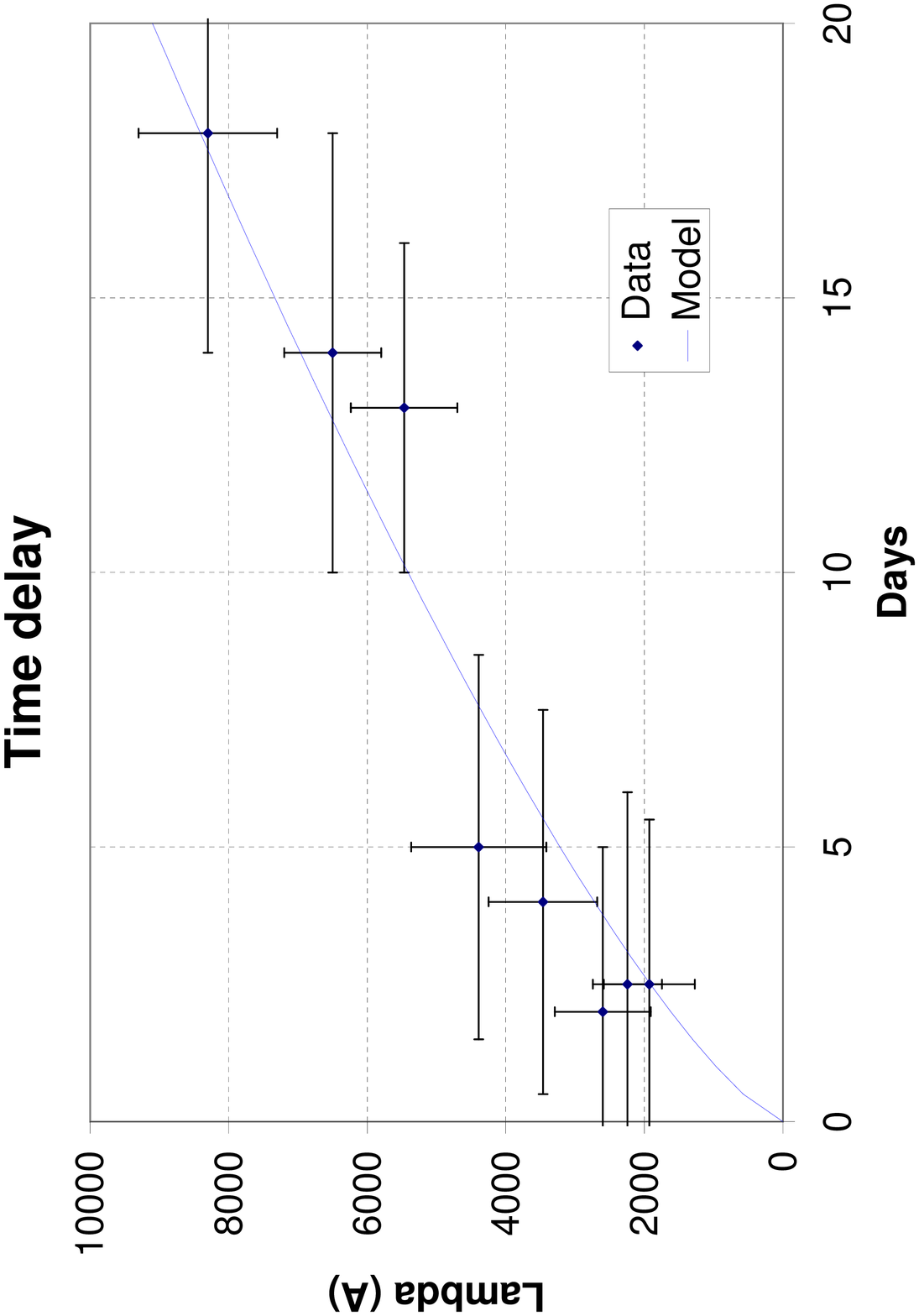} 

\caption[]{Time lags versus the wavelength. The wavelength uncertainties are due to the filter transmission curves and the time ones -- 
to the ICCF peak widths (see the text). Data points correspond to UVW2, UVM2, UVW1, U, B, V, R and I filters, from bottom to top, respectively.
For I-band, the first positive maximum is plotted, however a stronger maximum at $\tau \simeq -2$ days is evident from Figure 6, 
possibly indicating that the reprocessing perhaps also plays a role, but is not the main factor, leading to a casual connection between
the light curves. A fit to the data ($\tau_{\lambda}\simeq 9 \lambda_{5000}^{4/3}$ [days]) is also shown, indicating that the 
delay scales as expected for a standard accretion disk. The data, however, are not stringent enough to rule out other possibilities, 
since different fits are also possible.}

\end{figure}

\subsection{X-ray Spectroscopy}

\begin{figure}

 \mbox{} \vspace{6.5cm}  \includegraphics{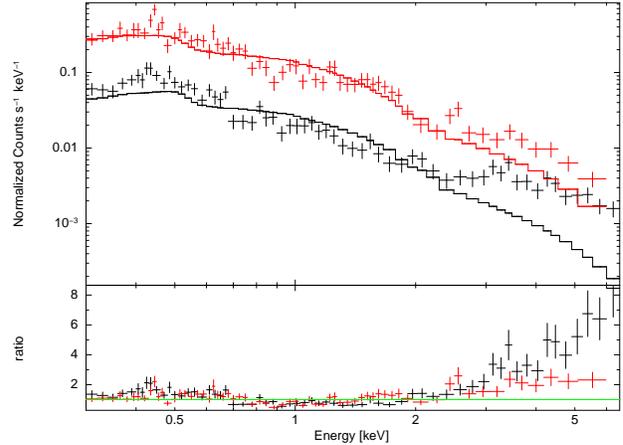} 

\caption[]{\label{xrt_xspec_plot}
\swift\ XRT low and high state spectra fitted by an absorbed single  power law model. 
The black spectrum is the low and the red the high state spectrum. }

\end{figure}

As described in Section 2.1, the data were combined to derive low, intermediate and high-state spectra 
of PG 1211+143. These data were first fitted by a single absorbed power law
model with the absorption column density fixed to the Galactic value (2.47$\times 10^{20}$
cm$^{-2}$; Kalberla et al. 2005). Table\,\ref{xrt_xspec} lists the spectral fit parameters.
Obviously a single power law model does not represent the observed spectrum. 
 Figure\,\ref{xrt_xspec_plot} displays this
fit simultaneously to the low and high state \swift\ spectra. 
As a comparison, Table\,\ref{xrt_xspec} also lists the results for the fits to the 2001 and 
2004 \xmm\ EPIC pn data.
As the next step we fitted the spectra with a broken power law model. Although this model significantly improves the 
fits and describes the spectra quite well, it is not a physical model.
Especially in the low-state the hard X-ray spectral slope appears to be very flat with \ax=--0.18. 
This behaviour 
is typical when the X-ray spectrum is either affected by partial covering absorption or
reflection (e.g. Turner \& Miller 2009, Grupe et al. 2008a).

Next the spectra were fitted with a power law model with a partial covering absorber.
These fits suggest a strong partial covering absorber in the low-state spectrum 
with an absorption column density of the order of 9$\times 10^{22}$ cm$^{-2}$ and a covering
fraction of 95\%. During the intermediate state the column density of the absorber decreases to 8$\times 10^{22}$ cm$^{-2}$
with a covering fraction of 93\% and drops down to $N_{\rm H, pc}=3.5\times 10^{22}$ cm$^{-2}$ and $f_{\rm pc}$=78\%
during the high state. In order to check whether the data can be self-consistently fit, we fitted the low and high-state
\swift\ data simultaneously in {\it XSPEC}. In this case we tied the column densities of the partial covering absorber and
the X-ray spectral slopes of both spectra but left the covering fractions and the normalisations free. Here we found an
absorption column density of $N_{\rm H,pc}=8.1\times10^{22}$ cm$^{-2}$ with covering fractions of 91\% and 86\% for the
low and high states, respectively.  
In all cases, \swift\ as well as \xmm,  
the X-ray spectral slope remains around \ax=2.1 and does not show any significant
changes within the errors. The partial covering absorber model can explain the variability seen in X-rays in PG 1211+143. The
results obtained from the \swift\ data during the high state are consistent with those derived from the 2001 and 2004
\xmm\ EPIC pn data.

\begin{figure}

 \mbox{} \vspace{6.5cm}  \includegraphics{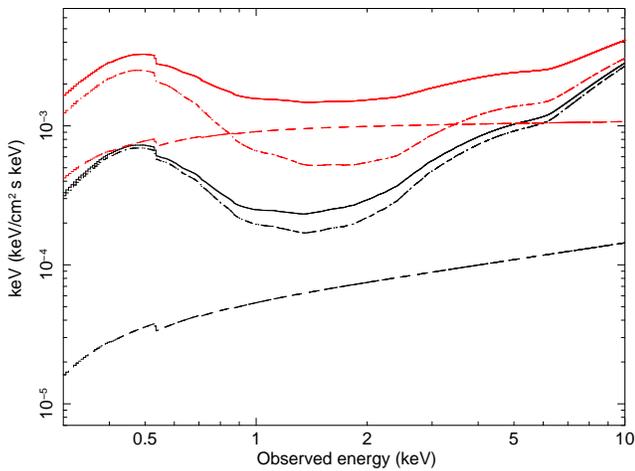} 

\caption[]{\label{pg1211_refl_mod}
Blurred reflection model applied to the low and high state XRT spectra of PG 1211+143 with the black
and red (the upper three) lines showing the models for the low and high state data, respectively.}

\end{figure}

\begin{figure}

 \mbox{} \vspace{6.5cm}  \includegraphics{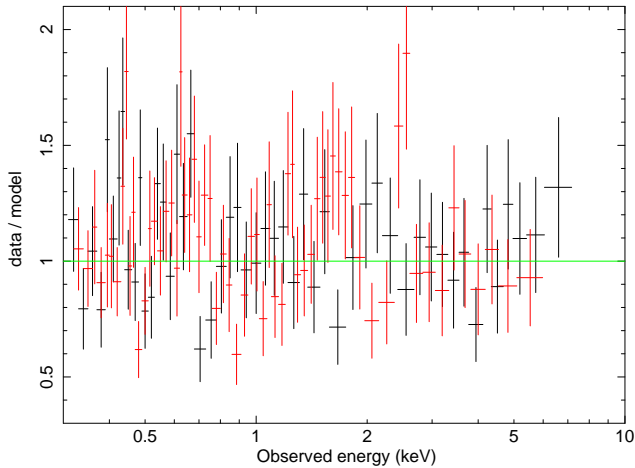} 

\caption[]{\label{pg1211_refl_ratio}
Ratio plot between the reflection model shown in Figure\,\ref{pg1211_refl_mod} and the data with the low
state data shown in black and the high state data in red.}

\end{figure}

The spectra were also fitted with a blurred reflection model
(Ross \& Fabian 2005).  Such models, where the primary emission (i.e. the
power law component) illuminates the accretion disk producing a reflection
spectrum that is blurred by Doppler and relativistic effects close to the
black hole (e.g. Fabian et al. 1989) have been successfully applied to several NLS1 X-ray spectra (e.g.
Fabian et al. 2004; Gallo et al. 2007a, 2007b; Grupe et al. 2008a; Larsson et al.
2008). As shown in Figures\,\ref{pg1211_refl_mod} and \ref{pg1211_refl_ratio},
the reflection model broadly describes the high- and low-flux states of PG~1211+143.
In the simplest case, the blurring parameters and disk ionisation are linked
between the two epochs.  The disk inclination ($i$) and outer radius
($R_{\rm out}$) are fixed to 30$^{\circ}$ and $400 R_{\rm g}$, respectively; $R_{\rm g}$ is the Schwarzschild radius.  
The continuum shape ($\Gamma$) and normalisations of the reflection and power law were free to vary
independently at each epoch.  The resulting fit is reasonable
($\chi^{2}_{\nu}/dof = 1.30/105$), considering the obvious
over-simplification of our model.
The inner disk radius and emissivity index were found to be $R_{\rm in}=1.76^{+0.27}_{-0.35} R_{\rm g}$ and
$q=5.57^{+0.54}_{-0.71}$, respectively.  The disk ionisation was $\xi = 116 \pm 10$.  The intrinsic
power law shape was significantly harder during the low-flux state,
$\alpha_{\rm x,low}=0.55^{+0.09}_{-0.06}$, compared to $\alpha_{\rm x,high}=0.96^{+0.13}_{-0.10}$
during the high-state.  The primary difference between the high and low state is the
relative contribution of the power law component to the total 0.3--10 keV flux, being approximately
0.42 and 0.12, respectively.  During the low-flux state the reflection component dominates
the spectrum.

\begin{table*}
  \centering
  \caption{\label{xrt_xspec} Spectral analysis of the X-ray low and high states of
  PG 1211+143 with the Galactic absorption column density set to 2.47$\times 10^{20}$
  cm$^{-2}$ given by Kalberla et al. (2005). The absorption column density of the
  partial covering absorber $N_{\rm H, pc}$ is given in units of $10^{22}$
  cm$^{-2}$. $f_{\rm pc}$ denotes the covering fraction of the partial covering
  absorber. The observed 0.2--2.0 and 2--10 keV fluxes $F_{\rm 0.2-2.0 keV}$ and $F_{\rm 2-10 keV}$ are given in units of
  $10^{-12}$ ergs s$^{-1}$ cm$^{-2}$.
 }
  \begin{tabular}{lcccccccc}
  \hline
  \hline

Model & $\alpha_{\rm x,soft}$ & $E_{\rm Break}$ & $\alpha_{\rm x,hard}$ & $N_{\rm H,pc}$ & $f_{\rm pc}$ & $\chi^2/\nu$ 
& $F_{\rm 0.2-2.0 keV}$ & $F_{\rm 2-10 keV}$ \\ 
\hline

\multicolumn{9}{c}{Low state} \\

\hline
Powl & 1.73\plm0.10 & --- & --- & --- & --- & 312/60 & 1.59\plm0.10 & 0.39\plm0.02 \\
Bknpo & 2.29$^{+0.14}_{-0.13}$ & 1.42$^{+0.13}_{-0.11}$ & --0.18$^{+0.17}_{-0.18}$ & --- & --- & 68/58 &
1.81\plm0.16 & 2.45\plm0.21 \\
Zpcfabs * powl & 2.18$^{+0.10}_{-0.12}$ & --- & --- & 9.45$^{+1.87}_{-1.62}$ & 0.95\plm0.02 & 82/58 
& 1.76\plm0.26 & 1.54\plm0.23 \\
\hline
\multicolumn{9}{c}{Intermediate state} \\
\hline
Powl & 1.71\plm0.15 & --- & --- & --- & ---  & 103/26 & 3.65\plm0.36 & 0.92\plm0.09 \\
Bknpo & 2.44\plm0.26 & 1.15$^{+0.27}_{-0.14}$ & 0.31$^{+0.26}_{-0.34}$ & --- & --- & 30/24 & 4.34\plm0.76 & 3.85\plm0.67
\\
Zpcfabs * powl & 2.22\plm0.20 & --- & --- & 7.88$^{+3.21}_{-2.33}$ &  0.93$^{+0.03}_{-0.04}$ &
  35/24 & 4.07\plm1.47 & 2.84\plm1.04 \\
\hline
\multicolumn{9}{c}{High state} \\
\hline
Powl & 1.47\plm0.06 & --- & --- & --- & --- & 168/62 & 7.04\plm0.29 & 2.76\plm0.12 \\
Bknpo & 2.05$^{+0.16}_{-0.15}$ & 1.04$^{+0.13}_{-0.12}$ & 0.84\plm0.14 & --- & --- & 91/60  &
7.79\plm0.78 & 5.06\plm0.51 \\
Zpcfabs * powl & 1.98\plm0.14 & --- & --- & 3.48$^{+1.99}_{-1.02}$ & 0.78$^{+0.05}_{-0.07}$ & 101/60 &
7.50\plm1.76 & 4.02\plm0.94 \\
\hline
\multicolumn{9}{c}{High and Low state simultaneously} \\
\hline
Powl & 1.56\plm0.05 & --- & --- & --- & --- & 488/123 & 1.49/7.24 & 0.50/2.41 \\
Bknpo & 2.14\plm0.10 & 1.30\plm0.12 & 0.34$^{+0.12}_{-0.12}$ & --- & --- & 215/121 & 1.79/7.65 & 1.62/6.92 \\
Zpcfabs * Powl & 2.00\plm0.09 & --- & --- &  8.11$^{+1.64}_{-1.33}$ & 0.91\plm0.02/0.86\plm0.03 & 209/120 & 1.65/7.68 &
1.53/4.52 \\
\hline
\multicolumn{9}{c}{XMM-Newton 2001} \\
\hline
Powl & 1.97\plm0.02 & --- & --- & --- & --- & 13092/994 & 5.56\plm0.03 & 0.86\plm0.01 \\
Bknpo & 2.20\plm0.01 & 1.28\plm0.05 & 0.71\plm0.05 & --- & --- & 4860/992 & 5.57\plm0.02 & 3.04\plm0.02 \\
Zpcfabs * powl & 2.15\plm0.01 & --- & --- & 6.41\plm0.15 & 0.87\plm0.02 & 5424/992 & 5.50\plm0.10 & 2.64\plm0.05 \\
\hline 
\multicolumn{9}{c}{XMM-Newton 2004} \\
\hline
Powl & 1.62\plm0.01 & --- & --- & --- & --- & 4122/940 & 5.63\plm0.05 & 1.06\plm0.02 \\
Bknpo & 1.75\plm0.01 & 1.58\plm0.05 & 0.85\plm0.03 & --- & --- & 1828/938 & 5.58\plm0.05 & 3.12\plm0.03 \\
Zpcfabs * powl & 1.73\plm0.01 & --- & --- & 8.73\plm0.34 & 0.70\plm0.02 & 1958/938 & 5.58\plm0.17 & 2.89\plm0.09 \\
\hline
\end{tabular}
\end{table*}

\subsection{Spectral energy distribution}

\begin{figure}

 \mbox{} \vspace{5.5cm}  \includegraphics{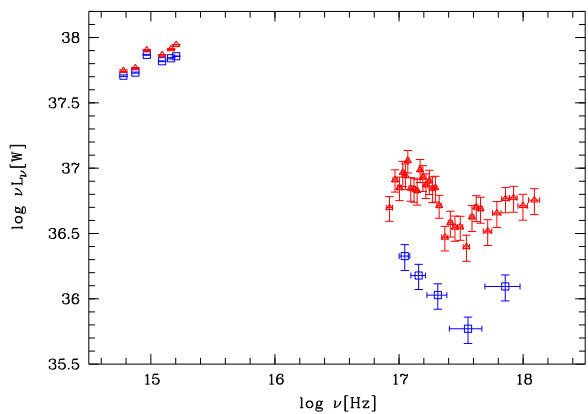} 

\caption[]{\label{pg1211_sed}
Spectral Energy Distributions of PG 1211+143 during the low and high states on 2007 April 17 and
22 (segments 017 and 018, respectively). The low state \swift\ UVOT and XRT data are displayed as
(blue) squares and the high state data as (red) triangles.}

\end{figure}

Figure\,\ref{pg1211_sed} displays the spectral energy distributions (SED) during the low state
observation on 2007 April 17 (blue squares) and the high state observation on April 22 (red
triangles). The optical/UV slope \auv\ slightly changes from $-$0.67\plm0.12 to
$-$0.56\plm0.10 between the low and high states. Most significant, however, is the change in the
optical-to-X-ray spectral slope \aox\footnote{The X-ray loudness is defined by Tananbaum et al. 
(1979) as \aox=--0.384 log($f_{\rm 2keV}/f_{2500\rm \AA}$).} from \aox=1.84 during the low state
to \aox=1.48 during the high state. This low-state \aox\ almost makes it an X-ray weak AGN according to the
definition by Brandt et al. (2000), who defines AGN with \aox$>$2.0 as X-ray weak. 
The luminosities in the Big-Blue-Bump are log $L_{\rm BBB}$=38.42 and 38.53 [W]
for the low and high states, respectively. These luminosities correspond to Eddington ratios of 
$L/L_{\rm Edd}$=0.26 and 0.33, respectively, assuming a mass of the central black hole of 9$\times
10^7$\msun\ (Vestergaard \& Peterson 2006).

\section{Discussion}

Based on the results from this study of the NLS1 PG 1211+143 we found that the  
short-time (and perhaps -- even the long-time) variations of the X-ray and optical/UV continuums do not seem to correlate well
(Smith \& Vaughan 2007, and the references within). 
While the X-ray continuum varied rapidly (more than 5 times during \swift\ monitoring campaign) 
with a general trend of brightness increase, 
the optical/UV continuum showed minimal changes with a general trend of brightness decrease. This is not unusual and is in fact reported
for other objects (e.g. NGC 5548, Uttley et al. 2003). Such a behaviour sets constraints on 
different reprocessing scenarios. We are going to discuss briefly several possibilities, assuming that the optical/UV emission is produced 
by a standard, thin accretion disk, the central black hole mass is $M_{\rm BH}\simeq 9 \times 10^{7}$ \msun\ 
and the accretion rate (in Eddington units) is $\dot m \simeq 0.3$ (Sect. 3.7). The accretion rate is also consistent within the errors 
with the one found by Kaspi et al. (2000) and Loska et al. (2004). 


\subsection{Cause of the X-ray weakness}

As shown in Figure\,\ref{longterm} over the last 20 years PG 1211+143 appears to be fainter in X-rays compared with the
X-ray observation during the 1980th. Especially during our \swift\ observations during the beginning of our monitoring
in March 2007, PG 1211+143 was found to be in an historical X-ray low state especially in the 0.2--2.0 keV band. 
The X-ray spectrum during the low state is somewhat similar to that found during the historical
low state in Mkn~335 (Grupe et al. 2007b, 2008a). Also, here the low state could be explained by a
strong partial covering absorber. Later monitoring with \swift\ suggests that the absorber has
disappeared again and that Mkn~335 is back in a high-state (Grupe et al. 2009, in prep.). 
Similarly, the low-state here was just a temporary event that lasted for a maximum of about a year. 
Partial covering absorbers, however, can last significantly longer. One example is the X-ray
transient NLS1 WPVS 007 (Grupe et al. 1995), which has developed strong broad absorption line features in
the UV (Leighly et al. 2009) and a strong partial covering absorber in X-rays (Grupe et al. 2008b).
It has been in an extreme low X-ray state for more than a decade (Grupe et al. 2007a, 2008b). 

Statistically we cannot distinguish between the partial covering absorber or the reflection models. Both models
result in similar $\chi^2/\nu$. Both models can also be fit self-consistently leaving the intrinsic X-ray spectrum fixed
and only affected by either partial covering absorption or reflection. 

  In the case of a partial covering absorber we can expect that the
 observed light is polarized like it has been seen in the NLS1 Mkn 1239 which is highly optically polarized
 and shows a strong partial covering absorber in X-rays (Grupe et al. 2004).
 However, both PG 1211+143 and Mkn 335 do not show any sign of optical continuum polarization
 (Berriman et al. 1999, Smith et al. 2002). Note, however, that all these polarimetry measurements were done when the
 objects were in their high-state. There are no polarimetry measurements (at least not to our knowledge) that were
 performed during their low states. Therefore the non polarization in the optical does not exclude the partial
 covering model.


\subsection{Compact central X-ray source or extended medium?}


It is commonly assumed that the most part of the AGN X-ray emission is produced close to the center, within the inner few tens $R_{\rm g}$. The X-rays in radio-quiet objects may come from the inner part of an accretion disk, perhaps operating there in a mode of a very hot, geometrically thick, low-efficient accretion flow (e.g. ADAF) or from an active corona, sandwiching the disk. In either case, when studying the X-ray irradiation of the \textit{optically} emitting outer parts of the disk (at $\sim 100-1000 R_{\rm g}$), the X-ray producing region could be considered as a point source, elevated slightly above the center of the disk and illuminating the periphery. Thus, a part of this highly variable by presumption X-ray emission could be reprocessed into optical/UV emission, which variations will lag behind the X-ray variations.

The temperature of a thin (Shakura-Sunyaev) accretion disk scales with the radial distance $r$ (in Schwarzschild radii) as 
$T(r) \simeq 6~10^{5}(\dot m /M_{\rm 8})^{1/4} r^{-3/4}$ [K] (Frank et al. 2002). 
If a point-like X-ray source, located at some distance $H$ above the disk, close to the center, irradiates the outer parts, it causes 
a temperature increase by a certain factor, but the radial temperature dependence happens to be the same (at least for $r>>H$). Since most of the 
visual/UV light is presumably due to viscous heating, not to irradiation, the temperature increase could be considered as a small 
addition to the usual disk temperature. Nevertheless, the X-ray variations should transform into \textit{some} optical/UV variations with a 
time delay, due to the light crossing time. If each disk ring emits mostly wavelengths close to the maximum of the 
\textit{Planck} curve for the corresponding temperature, one expects the time delays to scale with the wavelength as 
$\tau_{\rm \lambda} \simeq 5 (\dot m M_{\rm 8}^{2})^{1/3} \lambda_{\rm 5000}^{4/3}$ [days], which transforms to 
$\tau_{\rm \lambda} \simeq 5 \lambda_{\rm 5000}^{4/3}$ [days] for the accretion parameters accepted above.

The delay obtained from the fit in Sect. 3.5, however, appears to be $\sim$2 times longer than expected, yet consistent with the expected 
dependence of $\lambda^{4/3}$. Although different fits to the data are possible, due to the uncertainties, the general offset seems to 
indicate a time inconsistency. A possible explanation may be searched in the spatial location of the X-ray source. Even if the very 
center produces most of the X-ray emission, the outer disk may not "see" much of it. Instead, a large-scale back-scattering matter may be 
located at significant height above the disk, thus increasing the light-crossing time 2 -- 3 times (Loska et al. 2004; 
Czerny \& Janiuk 2007, see also their Fig. 6). In fact, the presence of such a back-scattering matter, in a form of a 
high-velocity outflow or a warm absorber is suggested by independent studies of this object (Pounds et al. 2003,  Pounds \& Page 2006).
One is to note however, that for a similar otherwise object -- Mkn~335, the delays were found to be consistent with a direct 
irradiation from a compact central source (Czerny \& Janiuk 2007). 
On the other hand, two recent studies by Ar\'evalo et al. (2008) and Breedt et al. (2009) found significant correlation between the 
X-ray and optical bands at essentially zero time lag (yet, not necessarily inconsistent with the direct reprocessing, considering 
the uncertainties) for MR 2251--178 and Mkn 79, respectively. Finally, for a broad-line radio galaxy (3C 120), an otherwise different type 
of object, but with similar to PG~1211+143 black hole mass and accretion rate, Marshall et al. (2009) found a $\sim$28 days 
delay of R-band behind the X-rays, being longer than expected and similar to our findings. 
Since the results from these and other studies have not shown a systematic inter-band behaviour, one approach to resolve the 
problem could be to study separately different types of objects, grouped by their intrinsic characteristics, like central masses, 
accretion rates, line widths, X-ray and radio properties, etc. (e.g. Bachev, 2009), in order to reveal how the presence of 
disk reprocessing might be related to overall quasar appearance. 


On the other hand, instead from around the central black hole, the variable X-ray emission could come from many active regions 
(flares) in the corona, covering the optically emitting parts of the disk (Czerny et al., 2004). Such active regions 
(or hot spots) can irradiate the disk locally, producing almost instantaneous optical/UV continuum changes. 
The later can be significant enough to be observable under some conditions. The observed wavelength-dependent time delay however, 
seems to make this possibility unlikely.

\subsection{Why is the optical/UV continuum poorly responding to the X-ray changes?}


A few factors can contribute to the apparent lack of significant correlation between the optical/UV and X-ray continua.
First, the possibility that the X-ray emission is highly anisotropic (Papadakis et al. 2000), or the disk geometry is far from flat 
(bumpy surface, Cackett et al. 2007; or warps, Bachev 1999), leading only to a minor optical response to the huge otherwise X-ray changes, 
cannot entirely be ruled out. Yet, the observed $\sim$2 times longer lags than expected are difficult to explain in terms of such an 
assumption, since in the presence of a large-scale backscattering medium, needed to account for the extra light travel path, the unevenness 
of the surface should be of little significance.

Another possibility, which might explain the large X-ray variations and the absence of optical/UV such is the presence of absorbing 
matter along the line of sight. If located close enough to the center, it can partially obscure the compact X-ray source from the observer, 
but not too much from the larger, optically emitting part of the accretion disk. In such a case, indeed, the large X-ray variations 
observed would hardly be transferred into the optical bands. Unfortunately, the X-ray spectral fitting cannot distinguish well 
enough between reflection and absorption models, to be able to determine which one shapes the X-ray continuum most.


\subsection{X-ray emission -- leading or trailing?}

Taking into account the position of the highest maximum of the I-band ICCF (Figure 6), the I-band changes appear to 
lead the X-ray ones. One way to explain this result, if real at all, is invoking the synchrotron-self Compton (SSC) 
mechanism to account for part of the produced X-rays. SSC assumes that some of the near-IR photons might have a 
synchrotron origin, and could later be scattered by the same relativistic electrons to produce the X-ray flares, 
lagging behind the infrared. However, the available data set, based merely on the light curves information, cannot 
justify undisputedly such an explanation. Furthermore, no strong jet or significant radio emission is present in this 
object (where SSC is typically assumed to play a significant role). So, if SSC is to account for the delay of the X-ray 
behind the I-band, this process has to take place in a the base of a possible failed jet (Czerny et al., 2008, and 
the references therein) or in the central parts of the disk, where the disk could operate as a very hot flow, 
and where hot electrons and perhaps strong magnetic fields could be present.


\section{Summary}

In this paper we presented the results of a continuum (X-ray -- to optical I-band) monitoring campaign of PG 1211+143, 
performed with \swift\ and ground-based observatories. The main results are summarized below:

\begin{enumerate}

\item 
In spite of being in a very low X-ray state, the quasar PG 1211+143 showed significant X-ray variations (up to 5 times) 
on daily basis, with only minor optical/UV flux changes. This behaviour indicates that a rather small amount of the hard 
radiation is reprocessed into longer wavelengths. Since both, reflection and absorption models fit equally well the 
X-ray spectrum, we are unable to determine the exact cause of the X-ray weakness of PG 1211+143 during its 2007 minimum. 

\item 
Interband cross-correlation functions suggest that a wavelength-dependent time delay between the X-ray and the 
optical/UV bands is present, indicating that at least a part of the X-rays is reprocessed into longer wavelengths.

\item 
Although the $\tau - \lambda$ dependence followed the general trend expected for a thin accretion disk 
(i.e. $\tau_{\rm \lambda} \sim \lambda^{4/3}$), the delays are $\sim$2 times longer, implying the possible 
existence of a large-scale back-scattering matter above the disk (wind/warm absorber), rather than a central 
point-like X-ray source, directly irradiating the disk.

\item
Even if the object is radio-quiet, with no strong jet known, we found indications that the SSC 
mechanism may play some role in the X-ray production.

\end{enumerate}

\section*{Acknowledgments}

We are grateful to Neil Gehrels for approving our ToO request and the \swift\ team for 
performing the ToO observations of PG 1211+143. We would also like to thank the anonymous referee 
for his/her helpful comments and suggestions which significantly improved this paper.
This research has made use of the NASA/IPAC 
Extragalactic Database (NED) which is operated by the Jet Propulsion Laboratory,
Caltech, under contract with the National Aeronautics and Space
Administration. \swift\ is supported at PSU by NASA contract NAS5-00136.
This research was supported by NASA contract NNX07AH67G (D.G.).


\newpage

\appendix

\section{Log of observations}

\begin{table*}
  \centering
  \caption{\label{obs_log} Summary of the \swift\ observations of PG 1211+143.
  Start and end times $T_{\rm start}$ and $T_{\rm end}$ of the observations are
  given in UT and all exposure times are given in units of s.
  }
  \begin{tabular}{ccccrrrrrrr}
  \hline
  \hline

Segment &  $T_{\rm start}$ & $T_{\rm end}$ & JD-2454000 & $T_{\rm XRT}$ & $T_{\rm V}$
&  $T_{\rm B}$ & $T_{\rm U}$ & $T_{\rm W1}$ & $T_{\rm M2}$ & $T_{\rm W2}$ \\
\hline
001 & 2007 Mar 08 00:48 & 2007 Mar 08 02:35 & 167 & 1111 & --- & --- & --- & --- & --- & --- \\
002 & 2007 Mar 09 01:03 & 2007 Mar 09 02:54 & 168 & 1781 & 148 & 148 & 148 & 298 & 373 & 596 \\
003 & 2007 Mar 10 17:20 & 2007 Mar 10 22:17 & 169 & 1695 & 154 & 154 & 154 & 309 & 203 & 621 \\
004 & 2007 Mar 11 19:03 & 2007 Mar 11 23:59 & 170 & 1711 & 154 & 154 & 154 & 310 & 217 & 621 \\
005 & 2007 Mar 12 15:55 & 2007 Mar 12 20:52 & 171 & 1666 & 149 & 149 & 149 & 300 & 219 & 601 \\
006 & 2007 Mar 13 20:50 & 2007 Mar 13 22:34 & 172 &  844 &  74 &  74 &  74 & 149 & 117 & 301 \\
007 & 2007 Mar 14 09:10 & 2007 Mar 14 12:34 & 173 & 1995 & 168 & 168 & 168 & 336 & 403 & 670 \\
008 & 2007 Mar 15 11:07 & 2007 Mar 15 13:02 & 174 & 1769 & 145 & 145 & 145 & 292 & 399 & 583 \\
009 & 2007 Mar 16 00:02 & 2007 Mar 16 17:54 & 175 & 1566 & 142 & 145 & 145 & 295 &  82 & 593 \\
010 & 2007 Mar 17 08:21 & 2007 Mar 17 19:12 & 176 & 2689 & 208 & 231 & 231 & 467 & 478 & 932 \\
011 & 2007 Mar 18 10:05 & 2007 Mar 18 16:39 & 177 & 1738 & 149 & 149 & 149 & 300 & 292 & 601 \\
012 & 2007 Mar 19 10:11 & 2007 Mar 19 16:45 & 178 & 1683 & 144 & 144 & 144 & 292 & 282 & 582 \\
013 & 2007 Mar 26 07:09 & 2007 Mar 26 08:28 & 185 & 2006 & 166 & 166 & 166 & 331 & 444 & 664 \\
014 & 2007 Apr 02 20:57 & 2007 Apr 02 22:36 & 192 & 2450 & 201 & 201 & 201 & 403 & 559 & 808 \\
016 & 2007 Apr 11 09:12 & 2007 Apr 11 14:09 & 201 & 1756 & 157 & 157 & 157 & 318 & 233 & 633 \\
017 & 2007 Apr 17 00:13 & 2007 Apr 17 06:46 & 207 & 2043 & 168 & 168 & 168 & 338 & 417 & 678 \\
018 & 2007 Apr 22 11:53 & 2007 Apr 22 15:18 & 212 & 2063 & 171 & 171 & 171 & 342 & 434 & 684 \\
019 & 2007 Apr 30 04:44 & 2007 Apr 30 06:32 & 220 & 1356 & 114 & 114 & 114 & 226 & 289 & 454 \\
021 & 2007 May 09 15:13 & 2007 May 09 17:04 & 229 & 1398 &  70 & 139 & 139 & 277 & 183 & 539 \\
022 & 2007 May 14 04:04 & 2007 May 14 05:59 & 234 & 2257 & 185 & 186 & 186 & 372 & 518 & 744 \\
023 & 2007 May 20 01:28 & 2007 May 20 03:12 & 240 & 857 & --- & 176 & 176 & 354 & --- & 120 \\
024 & 2008 Feb 17 07:56 & 2008 Feb 17 07:59 & 513 & 188 & ... & ... & ... & ... & ... & ... \\
\hline
\hline
\end{tabular}
\end{table*}

\begin{table*}
  \centering
  \caption{\label{xrt_uvot_res} XRT Count rates and hardness ratios and UVOT magnitudes of PG 1211+143. The
  UVOT magnitudes were corrected for Galactic reddening ($E_{\rm B-V}$=0.035; Schlegel et al. 1998). 
  }
  \begin{tabular}{ccccccccc}
  \hline
  \hline

Segment &  CR & HR & V & B & U & UV W1 & UV M2 & UV W2 \\

\hline
001 & 0.050\plm0.007 & --0.19\plm0.13 & ... & ... & ... & ... & ... & ... \\
002 & 0.057\plm0.006 & +0.02\plm0.11  & 14.32\plm0.02 & 14.59\plm0.01 & 13.33\plm0.01 & 13.31\plm0.01 & 13.16\plm0.01 & 13.18\plm0.01 \\
003 & 0.049\plm0.006 & --0.27\plm0.12 & 14.28\plm0.02 & 14.56\plm0.01 & 13.31\plm0.01 & 13.25\plm0.01 & 13.17\plm0.01 & 13.12\plm0.01 \\
004 & 0.085\plm0.008 & --0.14\plm0.09 & 14.26\plm0.02 & 14.58\plm0.01 & 13.29\plm0.01 & 13.24\plm0.01 & 13.15\plm0.01 & 13.12\plm0.01 \\
005 & 0.119\plm0.009 & --0.21\plm0.08 & 14.26\plm0.02 & 14.56\plm0.01 & 13.31\plm0.01 & 13.26\plm0.01 & 13.08\plm0.01 & 13.11\plm0.01 \\
006 & 0.059\plm0.009 & --0.17\plm0.15 & 14.29\plm0.02 & 14.52\plm0.01 & 13.33\plm0.01 & 13.27\plm0.01 & 13.07\plm0.03 & 13.11\plm0.01 \\
007 & 0.071\plm0.006 & --0.09\plm0.08 & 14.26\plm0.02 & 14.56\plm0.01 & 13.28\plm0.01 & 13.23\plm0.01 & 13.10\plm0.01 & 13.09\plm0.01 \\
008 & 0.060\plm0.006 & +0.08\plm0.06  & 14.26\plm0.02 & 14.58\plm0.01 & 13.37\plm0.01 & 13.28\plm0.01 & 13.17\plm0.01 & 13.18\plm0.01 \\
009 & 0.071\plm0.007 & --0.16\plm0.10 & 14.29\plm0.02 & 14.56\plm0.01 & 13.29\plm0.01 & 13.25\plm0.01 & 13.22\plm0.04 & 13.09\plm0.01 \\
010 & 0.059\plm0.005 & --0.10\plm0.09 & 14.27\plm0.02 & 14.54\plm0.01 & 13.31\plm0.01 & 13.26\plm0.01 & 13.12\plm0.01 & 13.12\plm0.01 \\
011 & 0.072\plm0.007 & --0.15\plm0.09 & 14.29\plm0.02 & 14.60\plm0.01 & 13.33\plm0.01 & 13.28\plm0.01 & 13.12\plm0.01 & 13.15\plm0.01 \\
012 & 0.045\plm0.006 & --0.05\plm0.12 & 14.29\plm0.02 & 14.60\plm0.01 & 13.38\plm0.01 & 13.27\plm0.01 & 13.14\plm0.01 & 13.13\plm0.01 \\
013 & 0.139\plm0.009 & --0.14\plm0.06 & 14.27\plm0.02 & 14.55\plm0.01 & 13.30\plm0.01 & 13.20\plm0.01 & 13.08\plm0.01 & 13.09\plm0.01 \\
014 & 0.171\plm0.008 & --0.22\plm0.05 & 14.27\plm0.02 & 14.57\plm0.01 & 13.30\plm0.01 & 13.24\plm0.01 & 13.10\plm0.01 & 13.11\plm0.01 \\
016 & 0.037\plm0.005 & --0.18\plm0.13 & 14.28\plm0.02 & 14.58\plm0.01 & 13.32\plm0.01 & 13.29\plm0.01 & 13.15\plm0.01 & 13.15\plm0.01 \\
017 & 0.060\plm0.006 & --0.17\plm0.09 & 14.36\plm0.02 & 14.68\plm0.01 & 13.43\plm0.01 & 13.37\plm0.01 & 13.28\plm0.01 & 13.32\plm0.01 \\
018 & 0.332\plm0.013 & --0.21\plm0.04 & 14.26\plm0.02 & 14.57\plm0.01 & 13.34\plm0.01 & 13.24\plm0.01 & 13.10\plm0.01 & 13.10\plm0.01 \\
019 & 0.095\plm0.009 & --0.16\plm0.09 & 14.33\plm0.02 & 14.58\plm0.01 & 13.30\plm0.01 & 13.24\plm0.01 & 13.11\plm0.01 & 13.16\plm0.01 \\
021 & 0.293\plm0.015 & --0.11\plm0.05 & 14.26\plm0.02 & 14.58\plm0.01 & 13.32\plm0.01 & 13.26\plm0.01 & 13.11\plm0.02 & 13.13\plm0.01 \\
022 & 0.294\plm0.012 & --0.15\plm0.04 & 14.27\plm0.02 & 14.60\plm0.01 & 13.39\plm0.01 & 13.26\plm0.01 & 13.17\plm0.01 & 13.17\plm0.01 \\
023 & 0.095\plm0.011 & --0.18\plm0.11 & ...           & 14.61\plm0.01 & 13.35\plm0.01 & 13.30\plm0.01 & ...           & 13.18\plm0.01 \\
024 & 0.375\plm0.045 & --0.10\plm0.12 & ... & ... & ... & ... & ... & ... \\
\hline
\hline
\end{tabular}
\end{table*}

\begin{table*}
\centering
\caption{Ground-based observations. All magnitudes are corrected for Galactic reddening ($E_{\rm B-V}$=0.035; Schlegel et al. 1998)}
\begin{tabular}{ccccccc}
\hline
\hline

JD (2454..)&	U	   &	 B	     &	 V		   &	   R	      &	   I	& Instr.     \\
\hline

116.52	& ...	    	   &14.79\plm	0.10 &  14.29\plm   0.03    & 13.96\plm   0.01 &  13.75 \plm   0.02& AOB \\
153.52	& ...	    	   &	...	     &  14.22\plm   0.04    &  ...	       &  ...		   & AOB \\
157.39	& ...	    	   &14.49\plm	0.04 &  14.24\plm   0.02    & 13.91\plm   0.02 &  13.72 \plm   0.02& RSh \\
171.63	& 13.28\plm   0.07 &14.50\plm	0.03 &  14.25\plm   0.02    & 13.91\plm   0.02 &  13.73 \plm   0.03& RSh \\
172.42	& 13.33\plm   0.05 &14.51\plm	0.02 &  14.27\plm   0.02    & 13.93\plm   0.02 &  13.74 \plm   0.02& RSh \\
174.34	& ...	    	   &14.58\plm	0.07 &  14.28\plm   0.02    & 13.94\plm   0.01 &  13.73 \plm   0.01& AOB \\
175.35	& ...	    	   &14.57\plm	0.07 &  14.28\plm   0.02    & 13.93\plm   0.01 &  13.74 \plm   0.01& AOB \\
176.31	& ...	    	   &14.47\plm	0.10 &  14.28\plm   0.02    & 13.94\plm   0.01 &  13.75 \plm   0.01& AOB \\
176.45	& 13.32\plm   0.03 &14.49\plm	0.03 &  14.26\plm   0.02    & 13.93\plm   0.02 &  13.72 \plm   0.02& RSh \\
177.32	& ...	    	   &	...	     &  14.28\plm   0.03    & 13.96\plm   0.02 &  13.73 \plm   0.03& AOB \\
178.32	& ...	    	   &14.55\plm	0.10 &  14.28\plm   0.02    & 13.92\plm   0.02 &  13.74 \plm   0.02& AOB \\
179.47	& ...	    	   &14.57\plm	0.10 &  14.26\plm   0.02    & 13.94\plm   0.02 &  13.74 \plm   0.02& AOB \\
199.49	& ...	    	   &14.50\plm	0.01 &  14.25\plm   0.01    & 13.98\plm   0.01 &  13.78 \plm   0.01& R2m \\
200.43	& ...	    	   &14.51\plm	0.01 &  14.23\plm   0.01    & 13.97\plm   0.01 &  13.76 \plm   0.01& R2m \\
201.44	& ...	    	   &14.56\plm	0.03 &  14.22\plm   0.03    & 13.94\plm   0.03 &  ...		   & R2m \\
201.45	& ...	    	   &14.57\plm	0.02 &  14.23\plm   0.01    & 13.97\plm   0.01 &  13.75 \plm   0.01& RSh \\
203.32	& ...	    	   &14.57\plm	0.01 &  14.23\plm   0.03    & 14.00\plm   0.01 &  13.78 \plm   0.01& RSh \\
204.34	& ...	    	   &14.63\plm	0.01 &  14.23\plm   0.01    & 13.97\plm   0.01 &  13.78 \plm   0.01& RSh \\
208.49	& 13.39\plm   0.03 &14.56\plm	0.02 &  14.28\plm   0.02    & 13.97\plm   0.02 &  13.77 \plm   0.02& RSh \\
210.41	& 13.35\plm   0.02 &14.56\plm	0.02 &  14.29\plm   0.02    & 13.97\plm   0.02 &  13.76 \plm   0.02& RSh \\
211.38	& 13.39\plm   0.03 &14.59\plm	0.02 &  14.31\plm   0.01    & 13.99\plm   0.02 &  13.77 \plm   0.02& RSh \\
213.29	& ...  	   &14.60\plm	0.05 &  14.31\plm   0.02    & 13.94\plm   0.01 &  13.76 \plm   0.02& AOB \\
217.29	& ...  	   &14.56\plm	0.05 &  14.36\plm   0.03    & 13.99\plm   0.02 &  13.77 \plm   0.02& AOB \\
230.35	& ...  	   &14.62\plm	0.05 &  14.28\plm   0.02    & 13.96\plm   0.01 &  13.76 \plm   0.01& AOB \\
231.30	& ...  	   &14.61\plm	0.05 &  14.27\plm   0.02    & 13.95\plm   0.01 &  13.74 \plm   0.01& AOB \\
232.31	& ...  	   &14.63\plm	0.06 &  14.29\plm   0.02    & 13.95\plm   0.02 &  13.74 \plm   0.02& AOB \\
233.42	& ...  	   &14.59\plm	0.05 &  14.38\plm   0.02    & 13.98\plm   0.01 &  13.78 \plm   0.01& AOB \\
234.30	& ...  	   &14.66\plm	0.06 &  14.23\plm   0.02    & 13.98\plm   0.01 &  13.77 \plm   0.01& AOB \\
236.32	& ...  	   &14.60\plm	0.07 &  14.31\plm   0.02    & 13.98\plm   0.01 &  13.77 \plm   0.01& AOB \\
238.37	& ...  	   &14.57\plm	0.01 &  14.23\plm   0.01    & 13.91\plm   0.01 &  ...		   & R2m \\
247.45	& ...	    	   &14.56\plm	0.02 &  14.29\plm   0.03    & 13.99\plm   0.03 &  13.81 \plm   0.03& RSh \\
260.39	& 13.35\plm   0.05 &14.60\plm	0.03 &  14.31\plm   0.02    & 14.00\plm   0.02 &  13.82 \plm   0.02& RSh \\
261.36	& ....  	   &14.62\plm   0.15 &  14.38\plm   0.02    & 14.02\plm   0.02 &  13.83 \plm   0.02& AOB \\
262.40	& ...  	   &14.65\plm   0.03 &  ...		    & 14.01\plm   0.02 &  13.79 \plm   0.03& RSh \\
263.36	& ...  	   &14.61\plm   0.03 &  14.33\plm   0.02    & 13.97\plm   0.02 &  13.81 \plm   0.03& RSh \\
265.35	& ...  	   &14.64\plm   0.10 &  14.39\plm   0.02    & 14.05\plm   0.02 &  13.87 \plm   0.02& AOB \\
265.42	& ...  	   &14.68\plm   0.02 &  14.35\plm   0.02    & 14.00\plm   0.03 &  13.84 \plm   0.02& RSh \\
266.40	& ...  	   &14.66\plm   0.02 &  14.33\plm   0.02    & 14.01\plm   0.02 &  13.81 \plm   0.02& RSh \\
289.31	& ...  	   &14.72\plm   0.13 &  14.36\plm   0.02    & 14.01\plm   0.02 &  13.84 \plm   0.02& AOB \\
290.35	& ...  	   &14.76\plm   0.13 &  14.35\plm   0.02    & 14.01\plm   0.02 &  13.85 \plm   0.03& AOB \\
291.34	& ...  	   &14.69\plm   0.02 &  14.32\plm   0.02    & 14.06\plm   0.01 &  13.89 \plm   0.01& RSh \\
292.31	& ...	    	   &14.68\plm	0.02 &  14.28\plm   0.02    & 14.03\plm   0.01 &  13.87 \plm   0.01& RSh \\
292.36	& ...	    	   &	...	     &  14.32\plm   0.06    & 13.95\plm   0.06 &  ...		   & AOB \\
293.32	& ...	    	   &14.67\plm	0.01 &  14.38\plm   0.02    & 14.04\plm   0.01 &  13.85 \plm   0.01& RSh \\
296.33	& ...	    	   &14.73\plm	0.03 &  14.38\plm   0.02    & 14.02\plm   0.01 &  ...		   & RSh \\
298.30	& ...	    	   &14.62\plm	0.03 &  14.37\plm   0.02    & 14.03\plm   0.03 &  13.85 \plm   0.03& RSh \\
299.30	& 13.51\plm   0.09 &14.66\plm	0.03 &  14.36\plm   0.02    & 14.02\plm   0.02 &  13.86 \plm   0.03& RSh \\
303.30	& ...	    	   &14.81\plm	0.02 &  14.38\plm   0.01    & 14.07\plm   0.01 &  13.89 \plm   0.02& R2m \\

\hline
\hline
\end{tabular}
\end{table*}

\bsp

\label{lastpage}


\begin{thebibliography}{}
\bibitem[]{} Abrassart A., Czerny B., 2000, A\&A 356, 475
\bibitem[]{}Ar\'evalo P., Uttley P., Kaspi S., et al., 2008, MNRAS, 389, 1479
\bibitem[Arnaud (1996)]{arnaud96} Arnaud K.~A., 1996, ASP
Conf.~Ser.~101: Astronomical Data Analysis Software and Systems V, 101, 17
\bibitem[]{} Alexander T., 1997, in "Astronomical Time Series", Eds. D. Maoz, A. Sternberg, and E.M. Leibowitz, (Dordrecht: Kluwer), p. 163
\bibitem[]{} Bachev R., 1999, A\&A, 348, 71
\bibitem[]{} Bachev R., Marziani P., Sulentic J. W., et al., 2004, ApJ 617, 171
\bibitem[]{} Bachev R., 2009, A\&A, 493, 907
\bibitem[Berriman et al.(1990)]{berriman90} Berriman G., Schmidt G.D., West S.C., 
\& Stockman H.S., ApJ Suppl, 74, 869 
\bibitem[]{} Boller T., Brandt W.N., \& Fink H.H., 1996, A\&A, 305, 53
\bibitem[]{} Boroson T.A., 2002, ApJ, 565, 78
\bibitem[Brandt et al.(1997)]{brandt97} Brandt, W.N., Mathus, S., \& Elvis, M., 1997, MNRAS, 285, L25
\bibitem[Brandt et al.(2000)]{brandt00} Brandt W.N., Laor A., \& Wills B.J., 2000, ApJ, 528, 637
\bibitem[]{} Breedt E., Ar\'evalo P., McHardy I. M., et al., 2009, MNRAS, 394, 427
\bibitem[Burrows et al. (2005)]{burrows04} Burrows D., et al., 2005, Space Science Reviews, 120, 165
\bibitem[]{} Cackett E. M., Horne K., Winkler H.,  2007, MNRAS, 380, 669
\bibitem[Chevallier et al.(2006)]{chevallier06} Chevallier L., Collin S., Dumont A.M., et al., 2006, A\&A, 449, 493 
\bibitem[]{} Chevalier C., Ilovaisky S. A.,  1991, A\&AS, 90, 225
\bibitem[]{} Czerny B., Rozanska A., Dovciak M., et al., 2004, A\&A, 420, 1
\bibitem[]{} Czerny B., Janiuk A.,  2007, A\&A, 464, 167
\bibitem[]{} Czerny B., Siemiginowska A., Janiuk A., Gupta A. C., 2008, MNRAS 386, 1557
\bibitem[Done \& Nayakshin (2007)]{done07} Done, C., \& Nayakshin, S., 2007, MNRAS, 377, L59
\bibitem[]{} Edelson R. A., Krolik J. H., 1988, ApJ, 333, 646
\bibitem[]{} Elvis M., Wilkes B.J. \& Tananbaum H., 1985, ApJ, 292, 357
\bibitem[Elvis (2000)]{elvis00} Elvis, M., 2000, ApJ, 545, 63
\bibitem[]{} Fabian A.C., Rees M.J., Stella L. \& White N.E., 1989, MNRAS, 238, 729 
\bibitem[]{} Fabian A.C., Miniutti G., Gallo L.C., et al.,  2004, MNRAS, 353, 1071
\bibitem[]{} Ferrarese L., Merritt D.,  2000, ApJ, 539, 9
\bibitem[Frank et al. (2002)]{fr} Frank J., King A., Raine D.: 2002, "Accretion Power in Astrophysics", Cambridge University 
\bibitem[Gallo (2006)]{gallo06} Gallo L.C., 2006, MNRAS, 368, 479
\bibitem[Gallo et al.(2997a)]{gallo07a} Gallo L.C., Brandt W.N., Constantini E., \& Fabian A.C., 2007a, MNRAS, 377, 1375
\bibitem[Gallo et al.(2997a)]{gallo07a} Gallo L.C., Brandt W.N., Constantini E., et al.,  2007b, MNRAS, 377, 391
\bibitem[]{} Gaskell C. M., Peterson B. M.,  1987, ApJS, 65, 1
\bibitem[]{Gas86} Gaskell C. M., Sparke L. S.,  1986, ApJ, 305, 175
\bibitem[]{} Gehrels N., Chincarini G., Giommi P., et al., 2004, ApJ 611, 1005
\bibitem[Grupe et al.(1995)]{grupe95} Grupe D., Beuermann K., Mannheim K., et al., 1995, A\&A, 300, L21
\bibitem[]{} Grupe D., Thomas H.-C. \& Beuermann, K., 2001, A\&A, 367, 470
\bibitem[]{} Grupe D., 2004, AJ, 127, 1799
\bibitem[]{} Grupe D., Schady P., Leighly K. M., et al., 2007a, AJ, 133, 1988 
\bibitem[Grupe et al.(2007b)]{grupe07} Grupe D., Komossa S., Gallo L.C., 2007b, ApJ, 668, L111
\bibitem[Grupe et al.(2008a)]{grupe08} Grupe D., Komossa S., Gallo, L.C., et al., 2008a, ApJ, 681, 982
\bibitem[Grupe et al.(2008b)]{grupe08b} Grupe D., Leighly K.M., Komossa S., 2008b, AJ, 136, 2343 
\bibitem[Grupe et al.(2004)]{grupe04} Grupe, D., Mathur, S., \& Komossa, S., 2004, AJ, 127, 3161
\bibitem[]{} Janiuk A., Czerny B., Madejski G. M., 2001, ApJ 557, 408
\bibitem[Hill et al. (2004)]{hill04} Hill J.E., et al., 2004, SPIE, 5165, 217
\bibitem[]{} Kalberla P. M. W., Burton W. B., Hartmann D., et al., 2005, A\&A, 440, 775
\bibitem[]{Kas00}	Kaspi S., Smith P. S., Netzer H., et al.,  2000, ApJ, 533, 631
\bibitem[]{} Kaspi S., Behar E., 2006, ApJ, 636, 674  
\bibitem[]{} Krolik J. H., 1999, "Active galactic nuclei : from the central black hole to the galactic environment", Princeton University Press.
\bibitem[]{} Landolt A. U., 1992, AJ, 104, 340
\bibitem[]{} Larsson J., Miniutti G., Fabian A.C., et al., 2008, MNRAS, 384, 1316
\bibitem[]{} Leighly K.M., Mushotzki R.F., Nandra K., \& Foster K., 1997, ApJ, 486, L25
\bibitem[]{} Leighly K.M., 1999a, ApJS, 125, 297
\bibitem[]{} Leighly K.M., 1999b, ApJS, 125, 317 
\bibitem[Leighly et al.(2009)]{leighly09} Leighly K.M., Hamann F., Casebeer D.A., Grupe D.,
2009, ApJ, submitted.
\bibitem[]{} Loska Z., Czerny B., Szczerba R., 2004, MNRAS, 355, 1080
\bibitem[]{} Magorrian J., Tremaine S., Richstone D., et al., 1998, AJ, 115, 2285
\bibitem[]{} Marshall K., Ryle W. T., Miller H. R., et al., 2009, ApJ, 696, 601
\bibitem[Miniutti \& Fabian(2004)]{miniutti04} Miniutti G., Fabian A.C., 2004, MNRAS, 349, 1435
\bibitem[]{} Osterbrock D. E., Pogge R. W., 1985, ApJ 297, 1660
\bibitem[]{} Papadakis I. E., Brinkmann W., Negoro H., et al., 2000, astro-ph/0012317
\bibitem[]{} Poole T., et al., 2008, MNRAS, 383, 627
\bibitem[]{} Pounds K. A., Reeves J. N., King A. R., et al., 2003, MNRAS, 345, 705
\bibitem[]{} Pounds K. A., Page K. L., 2006, MNRAS, 372, 1275
\bibitem[]{} Pounds K.A., Reeves J.N., 2007, MNRAS, 374, 823 
\bibitem[Roming et al. (2005)]{roming04} Roming P.W.A., et al., 2005, Space Science Reviews, 120, 95
\bibitem[Roming et al. (2009)]{roming09} Roming, P.W.A., et al., 2009, ApJ, 690, 163
\bibitem[Ross \& Fabian (2005)]{ross05} Ross R.R., Fabian A.C., 2005, MNRAS, 358, 211
\bibitem[Schlegel et al.(1998)]{sfd98} Schlegel D.~J., Finkbeiner D.~P., \& Davis M., 1998, ApJ, 500, 525
\bibitem[Smith et al(2002)]{smith02} Smith J.E., Young S., Robinson A., et al., 2002, MNRAS, 773, 798
\bibitem[]{} Smith R., Vaughan S., 2007, MNRAS, 375, 1479
\bibitem[]{} Sulentic J.W., Zwitter T., Marziani P., \& Dultzin-Hacyan D., 2000, ApJ, 536, L5 
\bibitem[Tananbaum et al. (1979)]{tananbaum79} Tananbaum H., et al., 1979, ApJ, 234, L9
\bibitem[]{} Turner T. J., Miller L., George I. M., Reeves J. N., 2006, A\&A, 445, 59
\bibitem[Turner \& Miller (2009)]{turner09} Turner T.J., \& Miller L., 2009, A\&A Review, 17, 47 
\bibitem[]{} Uttley P., Edelson R., McHardy I. M., et al., 2003, ApJ, 584, 53
\bibitem[Vestergaard \& Peterson(2006)]{vestergaard06} Vestergaard, M., \& Peterson, B.M., 2006, ApJ, 641,
689
\end{thebibliography}
\end{document}